\documentclass[12pt]{article}
\usepackage[utf8]{inputenc}
\usepackage{fancyhdr}
\usepackage{url}
\usepackage[shortcuts]{extdash}
\usepackage{graphicx}

 {\fancyhead{}
\fancyfoot[c]{\small{\rule{17.5cm}{1pt}}}}

\begin{document}
\title{\vspace{-2.5cm}
\begin{center}
\textbf{\small{ECIR WORKSHOP REPORT}}\\\vspace{-0.5cm} \rule{17.5cm}{1pt}
\end{center}
\vspace{1cm}\textbf{Report on the 7th International Workshop on Bibliometric-enhanced Information Retrieval (BIR 2018)}}

\author{Philipp Mayr \\
       GESIS -- Leibniz Institute for the Social Sciences, Germany \\
       \emph{philipp.mayr@gesis.org}
       \and
       Ingo Frommholz \\
       Institute for Research in Applicable Computing\\ University of Bedfordshire, Luton, UK\\
       \emph{ifrommholz@acm.org} \and
      Guillaume Cabanac \\
       University of Toulouse, Computer Science Department\\ IRIT UMR 5505, France\\
       \emph{guillaume.cabanac@univ-tlse3.fr} \\
       \date{}}

\maketitle \thispagestyle{fancy} 

\begin{abstract}	
The Bibliometric-enhanced Information Retrieval (BIR) workshop series has started at ECIR in 2014 and serves as the annual gathering of IR researchers who address various information-related tasks on scientific corpora and bibliometrics. We welcome contributions elaborating on dedicated IR systems, as well as studies revealing original characteristics on how scientific knowledge is created, communicated, and used. This report presents all accepted papers at the 7th BIR workshop at ECIR 2018 in Grenoble, France.
\end{abstract}

\section{Introduction}
The Bibliometric-enhanced Information Retrieval (BIR) workshop series has started at ECIR in 2014~\cite{Mayr2014} and serves as the annual gathering of IR researchers who address various information-related tasks on scientific corpora and bibliometrics~\cite{Mayr2015}. The workshop features original approaches to search, browse, and discover value-added knowledge from scientific documents and related information networks (e.g., terms, authors, institutions, references). 

The current incarnation is a continuation of the evolution of our workshop series. The first BIR workshops set the research agenda by introducing the workshop topics, illustrating state-of-the-art methods, reporting on current research problems, and brainstorming about common interests. For the fourth workshop, co-located with the ACM/IEEE-CS JCDL 2016, we broadened the workshop scope and interlinked the BIR workshop with the natural language processing (NLP) and computational linguistics field~\cite{Cabanac2016}. This joint activity has been continued in 2017 at SIGIR in the second BIRNDL workshop~\cite{mayr-SIGIRforum2017}. 

This 7th full-day BIR workshop at ECIR 2018\footnote{\url{http://bit.ly/bir2018}} aimed to foster a common ground for the incorporation of bibliometric-enhanced services (including text mining functionality) into scholarly search engine interfaces. In particular we addressed specific communities, as well as studies on large, cross-domain collections. This workshop strived to feature contributions from core bibliometricians and core IR specialists who already operate at the interface between scientometrics and IR. 

\section{Overview of the papers}
This year's workshop hosted two keynotes as well as a set of regular papers and two demos\footnote{Workshop proceedings are available at: \url{http://ceur-ws.org/Vol-2080/}}. The publications are briefly outlined in the following subsections.

\subsection{Keynotes}
\begin{figure}\centering
\includegraphics[width=\linewidth]{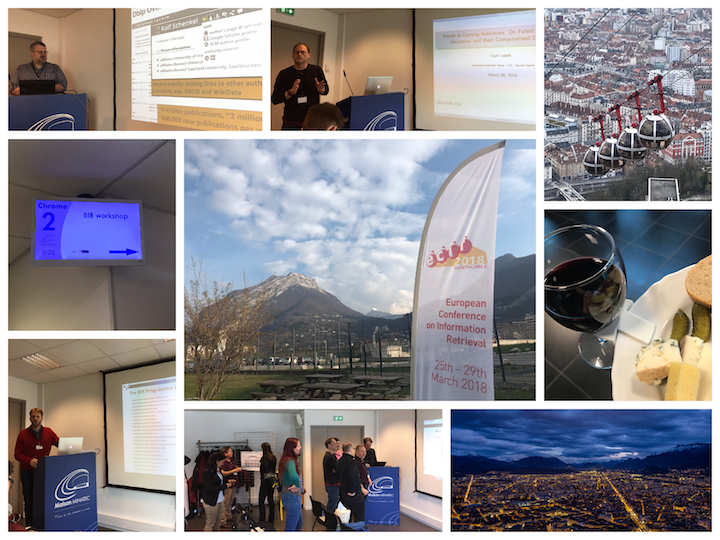}
\caption{A sense of the atmosphere at the BIR workshop.}\label{fig:collage}
\end{figure}

	This workshop featured two inspirational keynotes to kick-start thinking and discussion on the workshop topic. They were followed by paper presentations and demos (Fig.~\ref{fig:collage}) in a format that we found to be successful at previous BIR workshops.

	Cyril Labbé tackled a hot topic in his keynote titled ``Trends in gaming indicators: On failed attempts at deception and their computerised detection''~\cite{Labbe2018}.  He outlined various efforts to manipulate indicators by tricking the scientific community (e.g., by submitting automatically generated papers).  Other issues undermining the trust we place in peer-reviewed science were examined, such as data--results mismatch impeding the reproduction of results in cancer research. Labbé surveyed his recent work in these areas while reflecting on the potential of B+IR (bibliometrics and information retrieval) to address these critical issues.

	Ralf Schenkel presented in his keynote ``Integrating and exploiting  metadata sources in a bibliographic information system''~\cite{Schenkel2018} an in-depth summary of recent metadata activities in the computer science bibliography DBLP, which is maintained by Schloss Dagstuhl and University of Trier. He outlined procedures for monitoring, selecting, and prioritizing computer science venues for inclusion in the DBLP bibliography. A special focus was given to author disambiguation and utilization of citation data.

\subsection{Regular papers}
	Sarol, Liu, and Schneider proposed a citation and text-based publication retrieval framework~\cite{Sarol2018}. After the user provides some seed articles, the system collects papers connected by citations and applies a combination of citation- and text-based filtering methods. The framework is evaluated in a systematic reviewing task.

	Ollagnier, Fournier, and Bellot highlighted the central references of a paper based on the mining of its fulltext, quantifying the occurences of all in-text references~\cite{Ollagnier2018}. They benchmarked this approach compared to a system in production at OpenEdition,\footnote{\url{https://www.openedition.org}} and discuss the results in terms of enhanced relevance.

	In their article on query expansion, Rattinger, Le Goff, and Guetl combined word embeddings and co-authorship relations~\cite{Rattinger2018}. The set of documents used for pseudo-relevance feedback was enriched by similar documents from co-authors, applying a locally trained Word2Vec model. Adding similar documents from co-authors significantly improved the baseline.

	Bertin and Atanassova reported on the construction of the InTeReC dataset~\cite{Bertin2018}. Utilising different section types from PLOS articles, InTeReC consists of within-text references and their surrounding sentences. Additionally, verb phrases were extracted, providing an idea of the nature of the reference.

	Kacem and Mayr investigated the usage and influence of a specific search stratagem -- the Journal Run -- in an academic search engine log file \cite{Kacem2018}. They studied the frequency and stage of use of journal run as well as its impact on sessions. The authors found that the frequency of usage of the analyzed journals is not related to the impact factor within these sessions and that the size of the journal (Bradford Zones) has an insignificant correlation.

\subsection{Demo papers}
	Cataldi, Di Caron, and Schifanella designed the $d$-index to evaluate the degree of dependence of a researcher with respect to his/her co-authors over time. They implemented this indicator and demonstrate it online\footnote{\url{http://d-index.di.unito.it}} with DBLP as a bibliographic datasource~\cite{Cataldi2018}.

	The demo paper by Bessagnet presented a framework combining thematic, temporal, and spatial features of Twitter tweets in the field of Human and Social Sciences \cite{Bessagnet2018}. The author promoted 5~W dimensions (who, when, what, where, why) for the analysis of tweets. 

\section{Outlook}

While the past workshops laid the foundations for further work and also made the benefit of bringing information retrieval and bibliometrics together more explicit, there are still many challenges ahead. One of them is to provide infrastructures and testbeds for the evaluation of retrieval approaches that utilise bibliometrics and scientometrics. To this end, a focus of the proposed workshop and the discussion was on real experimentations~(including demos) and industrial participation. This line was started in a related workshop at JCDL~(BIRNDL~2016) and continued at SIGIR~(BIRNDL~2017), but with a focus on digital libraries and computational linguistics. Given the complex information needs scholars are usually facing, we emphasized on information retrieval and information seeking and searching aspects.

In July 2018 we will run the third iteration of the BIRNDL workshop\footnote{\url{http://wing.comp.nus.edu.sg/birndl-sigir2018/}} at the 41st SIGIR conference in Ann Arbor, MI, USA.  In conjunction with the BIRNDL workshop, the 4th CL-SciSumm Shared Task in Scientific Document Summarization\footnote{\url{http://wing.comp.nus.edu.sg/cl-scisumm2018/}} will be hold.

\section{Further Reading}
In 2015 we published a first special issue on ``Combining Bibliometrics and Information Retrieval'' in the \emph{Scientometrics} journal~\cite{Mayr2015}. 
A special issue on ``Bibliometrics, Information Retrieval and Natural Language Processing in Digital Libraries'' will appear in 2018 in the \emph{International Journal on Digital Libraries} \cite{BIRNDL-SP-IJDL}. 
Another special issue on ``Bibliometric-enhanced Information Retrieval and Scientometrics'' is in preparation for the \emph{Scientometrics} journal.  

Since 2016 we maintain the ``Bibliometric-enhanced-IR Bibliography''\footnote{\url{https://github.com/PhilippMayr/Bibliometric-enhanced-IR_Bibliography/}} that collects scientific papers which appear in collaboration with the BIR/BIRNDL organizers. We invite interested researchers to join this project and contribute related publications.

\section{Acknowledgement}
We wish to thank all those who have contributed to the workshop proceedings: all the contributing authors and the many reviewers who generously offered their time and expertise\footnote{The list of PC members can be found at \url{https://www.gesis.org/en/services/events/events-archive/conferences/ecir-workshops/ecir-workshop-2018/}}.

\bibliographystyle{splncs}
\bibliography{bibdb}
\end{document}